\documentclass[aps,superscriptaddress,floatfix,nofootinbib,preprintnumbers,eqsecnum,twocolumn]{revtex4-1}
\usepackage[utf8]{inputenc}
\usepackage{lipsum, graphicx}
\usepackage{amssymb,amsmath,slashed,multirow,mathtools}
\usepackage{comment,color,placeins,amsmath,amsfonts,enumitem}
\usepackage{cancel}
\usepackage[utf8]{inputenc}

\usepackage{subfigure}

\usepackage{hyperref}
\usepackage[usenames,dvipsnames]{xcolor}
\hypersetup
{
  colorlinks,
  linkcolor={blue!80!black},
  citecolor={blue!70!black},
  urlcolor={blue!70!black}
}

\makeatletter
\renewcommand*\env@matrix[1][\arraystretch]{%
  \edef\arraystretch{#1}%
  \hskip -\arraycolsep
  \let\@ifnextchar\new@ifnextchar
  \array{*\c@MaxMatrixCols c}}
\makeatother

\def\beq{\begin{equation}}
\def\eeq{\end{equation}}

\def\bea{\begin{eqnarray}}
\def\eea{\end{eqnarray}}

\newcommand{\met}{{\:/\!\!\!\! E_T}} 
 
\newcommand{\mptvec}{{\;/\!\!\!\! \vec{P}_T}}

\newcommand{\GeV}{\text{GeV}}

\begin{document}
\title{Demystifying Freeze-In Dark Matter at the LHC}
\author{Kyu Jung Bae}
\email{kyujung.bae@knu.ac.kr}
\affiliation{\mbox{Center for Theoretical Physics of the Universe, Institute for Basic Science, Daejeon 34126, Korea}}
\affiliation{Department of Physics, Kyungpook National University, Daegu 41566, Korea}
\author{Myeonghun Park}
\email{parc.seoultech@seoultech.ac.kr}
\affiliation{\mbox{Center for Theoretical Physics of the Universe, Institute for Basic Science, Daejeon 34126, Korea}}
\affiliation{Institute of Convergence Fundamental Studies and School of Liberal Arts,
Seoultech, 232 Gongneungro, Nowon-gu, Seoul, 01811, Korea}
\author{Mengchao Zhang}
\email{mczhang@jnu.edu.cn}
\affiliation{Department of Physics and Siyuan Laboratory, Jinan University, Guangzhou 510632, P.R. China}
\affiliation{\mbox{Center for Theoretical Physics of the Universe, Institute for Basic Science, Daejeon 34126, Korea}}


\begin{abstract}
Freeze-in mechanism provides robust dark matter production in the early universe. 
Due to its feeble interactions, freeze-in dark matter leaves signals at colliders which are often involved with 
long-lived particle decays and consequent displaced vertices (DV).
In this paper, we develop a method to read off mass spectrum of particles being involved in the DV events at the LHC.
We demonstrate that our method neatly works under a limited statistics, detector resolution and smearing effects. 
The signature of DV at the LHC can come from either highly suppressed phase-space or a feeble coupling of particle decay processes.
By measuring invisible particle mass spectrum, one can discriminate these two cases 
and thus extract information of dominant freeze-in processes in the early universe at the LHC. 
\end{abstract}

\maketitle

\FloatBarrier

\section{Introduction}
\label{sec:Introduction}

Particle dark matter (DM) is strongly supported by plethora of 
cosmological and astrophysical evidences (for reviews on particle dark matter see Ref.~\cite{Jungman:1995df, Bertone:2004pz, Bertone:2016nfn}).
Since there is no proper candidate of DM in the standard model (SM), it is considered as 
one of the most prominent clues to go beyond the SM.
The freeze-in mechanism provides a plausible answer to the origin of particle DM in the primeval Universe~\cite{Hall:2009bx},
and also an intriguing set of DM searches in broad area
({\it e.g.} non-thermal distribution of DM and its impacts\,\cite{Merle:2015oja,Konig:2016dzg,Roland:2016gli,Heeck:2017xbu,Bae:2017tqn,Bae:2017dpt}, DM direct detection\,\cite{Essig:2011nj,Hambye:2018dpi}, the large hadron collider (LHC) searches\,\cite{Barenboim:2014kka,Hessler:2016kwm,Calibbi:2018fqf,Belanger:2018sti})\footnote{For more broad aspects of freeze-in dark matter, readers can find reviews in\,\cite{Bernal:2017kxu}.}
although it generically contains tiny interactions with visible particles.


On the contrary to the conventional freeze-out dark matter,
the freeze-in dark matter never reaches the equilibration with the thermal plasma.
Due to its tiny interaction strength,
instead, DM annihilation processes are always inefficient and thus produced DM particles are accumulated
as the Universe expands.
Hence the correct DM abundance can be achieved despite of feeble interactions\,\cite{Hall:2009bx}.
Moreover, the production processes involve renormalizable couplings,
so they become more important at low temperature and consequently
dominant number of DM particles are produced near the threshold mass scale of the production processes.
Below the threshold scale, the production processes are highly suppressed by Boltzmann factor.
Therefore, DM abundance is independent of high temperature physics, {\it i.e.}, reheating after primordial inflation,
but is dependent on details of production processes which are {\it determined by DM particle interactions}.

The feeble nature of DM implies another observable footprints
if the DM mass is $\mathcal{O}(1)\,\textrm{keV}$.
Since DM particles are not equilibrated after being produced, their initial phase space distribution
does not change but is simply redshifted during the cosmic expansion.
As pointed out in the literature \cite{Bae:2017tqn,Bae:2017dpt}, phase space distribution of DM depends on
its production channels: 2-body decay, 3-body decay, $s$-channel and $t$-channel scattering, {\it etc}.
Such non-thermal distribution of DM particles impacts on small scale structures
and can be probed by the Ly-$\alpha$ forest 
observation~\cite{Viel:2005qj, Seljak:2006qw, Viel:2006kd, Viel:2007mv, Viel:2013apy, Baur:2015jsy, Irsic:2017ixq, Yeche:2017upn}.

Along with the cosmology, collider study also shows robust signatures of the freeze-in dark matter model. 
During the freeze-in processes, DM particles come out from thermal plasma of visible particles via feeble interactions between the dark sector and visible sector.
It implies the events at colliders where visible particles are produced in pairs 
and decay into DM particles plus visible particles. For example, higgsino pairs are produced and decay into axinos and higgs bosons.
In virtue of its tiny coupling, such decays show a signature of the displaced vertex (DV)
at hadron colliders.\footnote{We refer readers to Ref.\,\cite{Alimena:2019zri} for a recent review on long-lived particle searches at the LHC.}
The DV searches at the LHC have been considered 
to investigate various long-lived particles 
and have increased the sensitivity.
In addition, a possible improvement at the high-luminosity (HL) LHC with precision timing information
would suppress SM backgrounds leading to a better sensitivity of such long-lived particle searches\,\cite{Liu:2018wte,Kang:2019ukr}.

In order to examine genuine properties of the freeze-in dark matter model at hadron colliders,
we need to go one step further.
The DM abundance and distribution are determined by couplings and mass spectrum of the dark sector including the DM component.
Thus it is essential to extract the mass information from DV events at colliders.
In the case of prompt decay events, kinematic analyses require large number of events
to read off the end-point in the invariant mass distribution\,\cite{Hinchliffe:1996iu}.
As pointed out in the early study\,\cite{Park:2011vw,Cottin:2018hyf}, on the other hand,
DV renders mass reconstruction possible in an event-by-event basis,
so it allows us to extract mass spectrum of dark sector although
a handful of events are available.
In reality, however, kinematic reconstruction in an event-by-event basis highly depends on 
detector effects including smearing of final state visible particles.

In this paper, we present a collider method to reconstruct relevant mass spectrum with only conventional track and calorimeter information. To reduce uncertainties in mass measurement originating from detector smearing effects, we develop a simple ``filtering" algorithm, which systematically discards the outliers in mass measurement of DV events. 
For an illustration of our method, we consider pair production of mother particles ($F$) and their subsequent decays into
dark matter particles ($\chi$) and $Z$ bosons, where $Z$ boson decays into a lepton pair for a precise reconstruction.

This paper is organized as follows.
In Sec.\,\ref{sec:rev}, a brief review on freeze-in dark matter is given 
to argue why the long-lived particle at the LHC is appropriate to examine cosmological 
freeze-in dark matter scenarios.
In Sec.\,\ref{sec:kin}, we explain kinematic relations in events with DV to reconstruct 
masses of the decaying particle $F$ and dark matter particle $\chi$. In Sec.\,\ref{sec:filter}, we demonstrate how much a simple filtering algorithm enhances accuracy by removing events which have strong smearing effects from a detector.
Finally we present our result in mass reconstructions with various benchmark points at the HL-LHC.

\section{A Brief Review on Freeze-in Dark Matter}
\label{sec:rev}

In freeze-in dark matter models, DM particles have tiny {\it renormalizable} couplings with thermal plasma
by which the freeze-in processes are mediated.
In the following, we will explain the freeze-in mechanism with a toy example.

Suppose that DM particles are produced by decay of particle $A$ which is in thermal equilibrium, {\it i.e.} $A\to B+\chi$.
Here we assume $B$ is also in thermal equilibrium.
In this case, dark matter production rate is determined by decay width of $A$, $\Gamma_A$.
The dominant production occurs when the plasma temperature $T$ is around $m_A$.
For $T<m_A$, population of $A$ is highly suppressed by the Boltzmann factor, so the production process becomes ineffective.
One finds the yield of DM particles~\cite{Hall:2009bx,Chun:2011zd}, 
\begin{eqnarray}
Y_{\chi}&\approx& \frac{135 g}{4\pi^4g_*^{3/2}}\frac{M_P \Gamma_A}{m_A^2}\int^{\infty}_{0}z^3K_1(z)dz\nonumber\\
&\approx&10^{-3}\frac{M_P\Gamma_A}{m_A^2}\, ,
\end{eqnarray}
where $g$ is the degrees of freedom of $A$, $g_*$ is the effective degrees of freedom of thermal plasma $M_P$ is the reduced Planck mass, and $K_1(z)$ is the first modified Bessel function of the second kind.
In the second line, we have taken $g_*=100$.
For the decay of $A$ via renomarlizable coupling $\lambda$, the decay width is given by
\begin{equation}
\Gamma_A\sim\frac{1}{8\pi}\lambda^2m_A.
\label{eq:decay}
\end{equation}
The yield becomes
\begin{equation}
Y_{\chi}\sim 4\times10^{-5}\lambda^2\frac{M_P}{m_A},
\end{equation}
and the DM density is given by
\begin{eqnarray}
\Omega_{\chi}h^2&\sim& 2.8\times 10^{8}Y_{\chi}\left(\frac{m_{\chi}}{\text{GeV}}\right)\nonumber\\
&\sim&0.1\left(\frac{\lambda}{10^{-8}}\right)^2\left(\frac{200~\text{GeV}}{m_A}\right)
\left(\frac{m_{\chi}}{10~\text{keV}}\right).
\end{eqnarray}
Therefore, one can see that the freeze-in process provides the correct DM abundance in a wide rage of 
couplings and DM masses although the coupling is tiny.
For the coupling $\lambda\sim 10^{-8}$, DM mass is of order keV, so it can be warm dark matter (WDM) and may be probed by small scale observables~\cite{Merle:2015oja,Konig:2016dzg,Roland:2016gli,Heeck:2017xbu,Bae:2017tqn,Bae:2017dpt}.
For even smaller coupling, $\lambda\sim10^{-12}$, on the other hand, weak scale DM mass ($\sim$100 GeV) is possible 
so that DM can be cold dark matter (CDM).

In most of freeze-in DM models, scattering processes are also accompanied by the decay processes.
One can find a process like $A+C\to \chi+D$ ($t$-channel mediated by $B$) where $C$ and $D$ are also in thermal equilibrium.
The single processes are normally less effective than the decay processes due to the suppression in the kinematic phase factor. 
In some cases, nonetheless, the scattering process can be enhanced by large number of degrees of freedom of accompanying particles $C$ and $D$, so it can make sizable contributions to the DM production (see Ref.~\cite{Bae:2011jb,Bae:2011iw} for axino production cases).
In addition, there may exist decay processes with more than 2 final state particles.
However, this contribution is more suppressed than 2-body decays by the kinematic phase factor, so is typically subdominant.

In the case where decay and scattering processes are ``co-resident,'' more interestingly, mass difference between $A$ and $B$ can determine
the relative ratio of decay contribution and scattering contribution as well as the phase space distribution. 
In principle, if $m_A-m_B\ll m_A$, the kinetic energy of produced DM can be arbitrarily small.
Under these circumstances, decay process become very inefficient.
Nevertheless, the scattering process does not alter dramatically, so it dominates the phase space distribution and relic abundance~\cite{Bae:2017dpt}.
Extracting mass spectrum at collider experiments may enable us to infer which process is dominant in freeze-in DM production.

At colliders, direct production of $\chi$ is impossible since its coupling to SM particles are too small.
However, its mother particle (we named $A$ in the above example) can be produced in the collision experiments.
If $A$ is the lightest parity-odd particle in the visible sector ({\it e.g.}, higgsino in the minimal supersymmetric standard model), it can decays only to DM particle with decay width in Eq.~\eqref{eq:decay}.
The decay length at the collider is
\begin{equation}
\frac{1}{\Gamma_A}\sim (1.7~\text{ns}~\text{or}~50~\text{cm})\times \left(\frac{10^{-8}}{\lambda}\right)^2\left(\frac{200~\text{GeV}}{m_A}\right).
\end{equation}
For the freeze-in DM with $m_{\chi}\sim10$~keV and $\lambda\sim10^{-8}$,
DV length is of order $10$~cm, and thus this gives {\it nearly the best sensitivity} for DV searches at the LHC~\cite{Aaboud:2017iio} while it gives dominant dark matter abundance.

For the larger dark matter mass, $m\sim 100$~GeV, smaller coupling $\lambda\sim10^{-12}$ is required 
to obtain the correct relic abundance in standard cosmology.
If this is the case, the decay length is too long to be covered by DV searches.
Such region, instead, may be covered by an additional long-lived particle detector outside of the LHC detectors\,\cite{Curtin:2018mvb}.
However, even in the case of large dark matter mass, there are viable non-standard cosmological scenarios leading to the correct relic abundance.
If the dark matter mass is around $100$~GeV and coupling is of order $10^{-8}$, 
the freeze-in process overproduces DM particles.
In such a case, large entropy dilution of factor $10^7-10^8$ is necessary to suppress freeze-in processes properly.
In a scenario where the DM freeze-in occurs during an early matter dominated era ({\it i.e.}, $T_R\ll100$~GeV),
relatively large coupling $\lambda\sim10^{-8}$ still provides a viable DM scenario with the correct relic DM abundance~\cite{Co:2015pka}.
In a scenario in the fast-expanding Universe~\cite{DEramo:2017ecx}, a similar dilution effect is possible so that viable freeze-in DM model is
possible with $m_{\chi}\sim100$~GeV and $\lambda\sim10^{-8}$.
In another freeze-in DM model~\cite{Merle:2015oja,Konig:2016dzg}, DM particle can be produced by decay of frozen-out particles rather than directly produced from thermal plasma.
In such a case, DV searches at the LHC covers the freeze-in DM in indirect way by showing kinematic structure of particle decaying into the DM particle.

In summary, the current and future DV searches for decay length of order $10$~cm
will be good probes of freeze-in WDM region and part of heavy freeze-in DM models.
Once we observe an excess of DV signals above the expected background in the future,
analyses on mass spectrum by using the kinematic techniques will be essential to 
reflect DM production process in the early Universe.
In next section, we show kinematic relations to reconstruct mass of mother particle and dark matter particle.

\section{Kinematics of Displaced Vertex }
\label{sec:kin}

We consider a case where the LHC produces pair of unstable particles $(F_1, F_2)$. These particles will decay into dark matter $\chi$ and $Z$ boson. 
As we have not seen any hints of dark matter at the LHC by Run 2, we do not expect to have enough statistics even for a discovery by the end of the  HL-LHC. Thus we need to consider a method which provides $(m_F, m_\chi)$ in the event-by-event basis, not through the accumulated information. To measure $(m_F, m_\chi)$ with a single event, we need to reconstruct four-momentum of dark matter particles ($\chi_i$).
As the number of unknowns from $\chi_i$ is eight in total, we need to achieve the same number of constraints using kinematics. 

\begin{figure}[tb]
\centering{\includegraphics[width=0.4\textwidth]{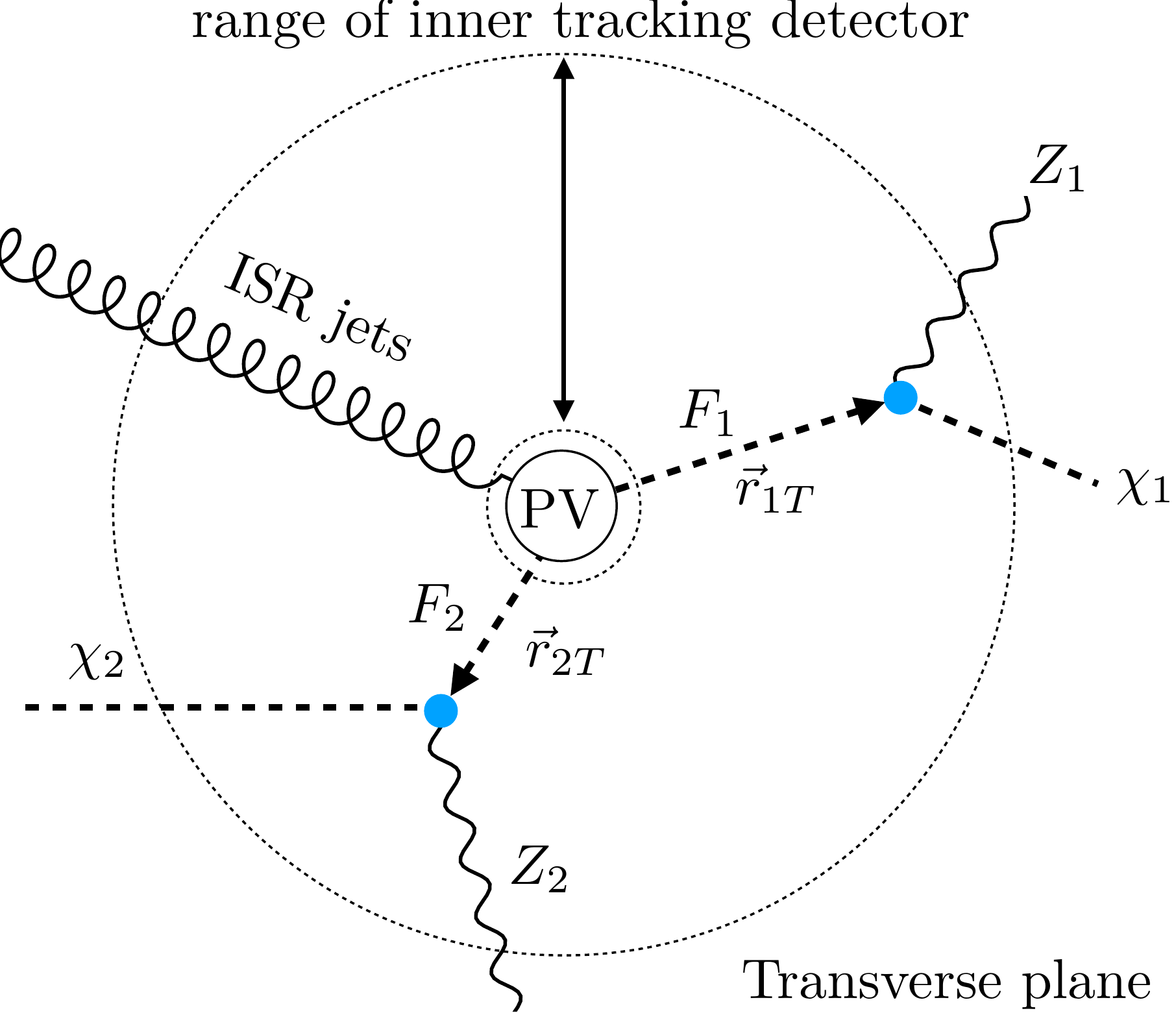}}
\caption{Positions of two DVs are denoted by $\vec{r}_1$ and $\vec{r}_2$ (dashed arrows).
Visible (reconstructable) particles are marked with solid lines while invisible particles with dashed ones.
Dotted circles represents inner and outer boundary of ``inner tracking detector(ID)" with cylindrical radius of $\mathcal{O}{(10)}-\mathcal{O}(10^3)$ mm. PV stands for primary vertex.
}
\label{fig:FFchichi}
\end{figure}

At the LHC detector, $F_i$ will leave a DV, denoted by $\vec r_i$, when it decays into $Z_i\,(\to \ell^+\ell^-)$ and $\chi_i$ as in  Fig.\,\ref{fig:FFchichi}. Due to the charge neutrality $F_i$, a three momentum vector $\vec p_{F_i}$ is proportional to $\vec r_i$. 
This provides two constraints for the direction of each $\vec p_{(F_i)}$, resulting in four constraints in total.  If we specify a direction of DV in a spherical coordinate  ($\hat{r}_i$: unit vector directing $\vec{r}_i$), 
\beq
\hat r_i = \left(\sin\theta_i \cos\phi_i,\, \sin \theta_i \sin \phi_i,\, \cos\theta_i \right)\,,
\eeq
we can express three momentum $\vec p_{(F_i)}$ in terms of DV position vector $\hat r_i$,
\beq
\vec p_{(F_i)} = |\vec p_{(F_i)}| \, \hat r_i\,.
\label{eq:DVP}
\eeq

In conventional searches for dark matter at the LHC, we utilize a momentum conservation in the transverse plane as we do not see the trace of dark matter directly. With this, we have two constraints:  
\beq
\mptvec = \sum_i \vec p_{T\,(\chi_i)} = -\left(\vec p_{T\,(\textrm{ISR})} + \sum_i \vec p_{T\, (Z_i)} \right)\,. 
\eeq
This can be simply translated into a condition for $\vec p_{(F_i)}$:
\beq
\sum_i \vec p_{T\,(F_i)} = -\vec p_{T\,(\textrm{ISR})}\,.
\label{eq:MET}
\eeq
The existence of ISR jets is important to identify dark matter at the LHC. For example, ISR jets are required for the most conventional dark matter searches\,\cite{Aaboud:2016tnv, CMS:2017tbk} 
and for utilizing information from a timing layer\,\cite{Liu:2018wte,Kang:2019ukr}. For the case of $\mathcal{O}(100-1000)\,\GeV$ mass scale of $F$, there are non-negligible number of events with $p_{T\,(\textrm{ISR})} \simeq \mathcal{O}(10-100)\,\GeV$\,\cite{Alwall:2007fs, Alwall:2009zu} and we can utilize this information to extract properties of dark matter at the LHC\,\cite{Feng:2005gj,Bae:2017ebe}. In our case, we use $p_{T\,(\textrm{ISR})}$ to reconstruct three-momenta $\vec p_{T\,(F_i)}$ by combining Eq.\,(\ref{eq:DVP}) and Eq.\,(\ref{eq:MET}),
\beq
\sum_i \left(|\vec p_{T\,(F_i)}| \,\hat r_{i\,T}  \right) = -\vec p_{T\,(\textrm{ISR})}\,.
\label{eq:METS}
\eeq
If we express above equation more specifically, it can be written as a $2\times 2$ matrix equation,
\bea
\begin{pmatrix} 
\sin\theta_1 \cos\phi_1 & \sin\theta_2 \cos\phi_2 \\
\sin\theta_1 \sin\phi_1 & \sin\theta_2 \sin\phi_2
\end{pmatrix} 
&&
\begin{pmatrix} 
|\vec p_{T\,(F_1)}| \\
|\vec p_{T\,(F_2)}|\\
\end{pmatrix} 
 \nonumber \\
= \,- &&
\begin{pmatrix} 
 |\vec p_{T\,(\textrm{ISR})}|\,\cos\phi \\
 |\vec p_{T\,(\textrm{ISR})}|\,\sin\phi
\end{pmatrix}\, , \,
\label{eq:MATRIX}
\eea
where $\phi$ is the azimuthal angle of $\vec p_{T\,(\textrm{ISR})}$ in a transverse plane. 
This provides a solution for $|\vec p_{T\,(F_i)}| $ in terms of directional angles $\left(\phi_i,\,\theta_i\right)$ of DVs,
\bea
&&|\vec p_{(F_1)}| = - \frac{  |\vec p_{T\,(\textrm{ISR})}|\, \sin(\phi-\phi_2)}{\sin\theta_1 \sin(\phi_1-\phi_2)} ,  \nonumber \\
&&|\vec p_{(F_2)}|  = \frac{ |\vec p_{T\,(\textrm{ISR})}|\, \sin(\phi-\phi_1)}{\sin\theta_2 \sin(\phi_1-\phi_2)} \, .
\label{Fsolution}
\eea
When ISR jet becomes soft, Eq.\,(\ref{eq:MATRIX}) turns into an indeterminate system. In the limit of $p_{T\,(\textrm{ISR})} \ll \sqrt{\hat s}$ where $\sqrt{\hat s}$ is the hard scale of an event, the difference $\phi_1-\phi_2$ can be approximated as
\beq
 \phi_1-\phi_2 \simeq \pi - \frac{2p_{T\,(\textrm{ISR})}}{\sqrt{\hat s}}\left(1-\frac{4 m_F^2}{\hat s}\right)^{-\frac{1}{2}} \to\,  \pi \, .
\eeq
As Eq.\,(\ref{eq:MATRIX}) has a factor $\sin\left(\phi_1-\phi_2\right)$ in its determinant, it is required to have non-negligible $p_{T\,(\textrm{ISR})}$ to have a numerically stable solution for $|\vec p_{(F_i)}|$. 

\begin{figure}[tb]
\centering{\includegraphics[width=0.4\textwidth]{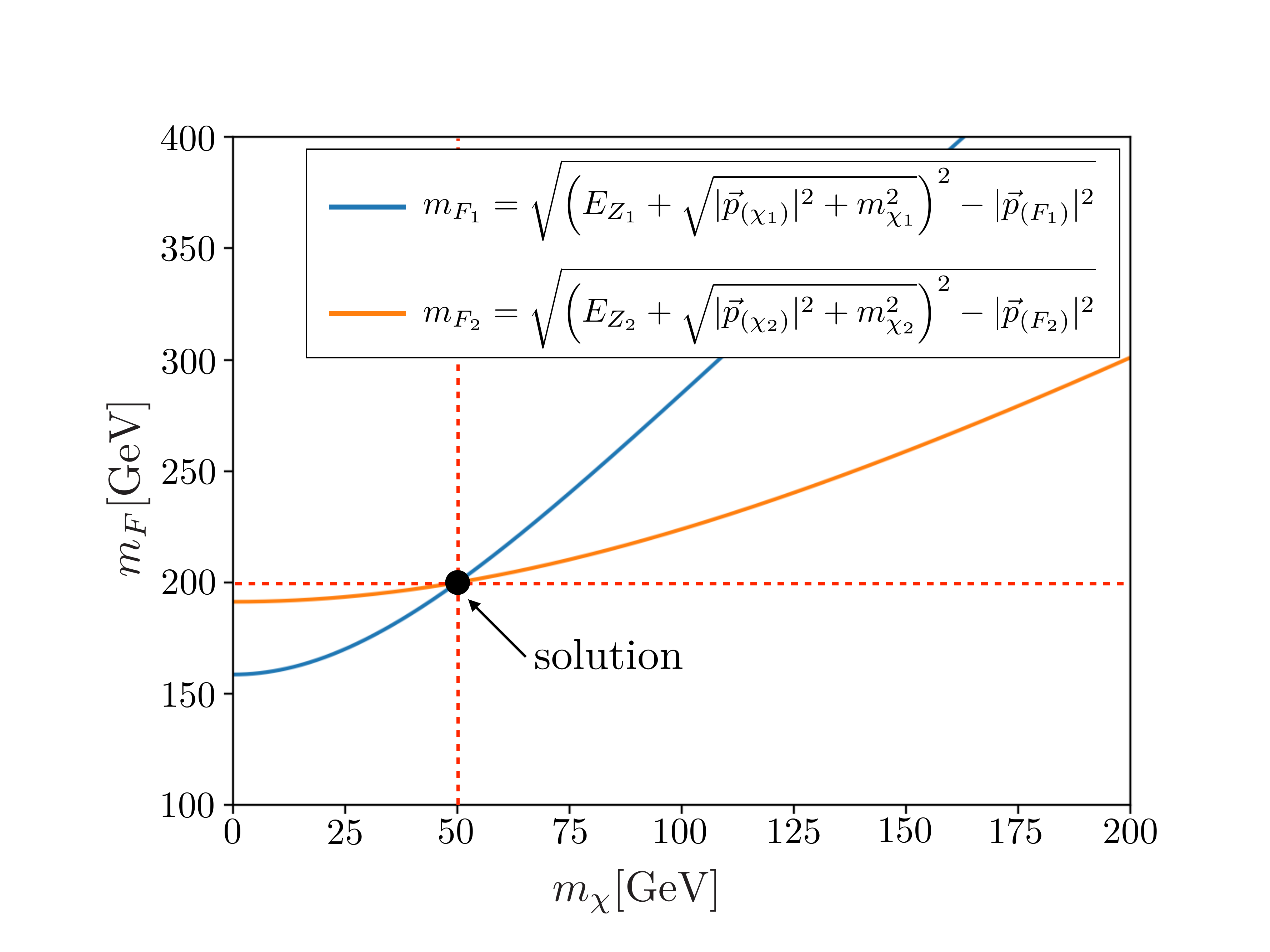}}
\caption{The functional dependence of $m_{F_i}$ on $m_{\chi_i}$ in a specific event is presented as blue and magenta lines according to Eq.\,(\ref{eq:func}). The event is generated at the parton level with a study point of $(m_F, \, m_\chi) = (200,50)\,\GeV$.
For $m_{F_1} = m_{F_2}$ and  $m_{\chi_1} = m_{\chi_2}$, the crossing point of two equations gives the solution for $m_F$ and $m_\chi$. 
}
\label{fig:solution}
\end{figure}

To extract information on masses of invisible particles, 
we convert solution of momenta into masses using the energy conservation as 
\beq
\sqrt{m_{F_i}^2+|\vec p_{(F_i)}|^2} = E_{Z_i} + \sqrt{m_{\chi_i}^2 + |\vec p_{(\chi_i)}|^2}\, .
\eeq
From this, we have a functional dependence of $m_{F_i}$ on $m_{\chi_i}$ as 
\beq
m_{F_i}= \sqrt{\left(E_{Z_i}+ \sqrt{|\vec p_{(\chi_i)}|^2+ m_{\chi_i}^2}\right)^2 - |\vec p_{(F_i)}|^2}\, .
\label{eq:func}
\eeq
Here the three-momentum of $\chi_i$ is given by $\vec p_{(\chi_i)} =  |\vec p_{(F_i)}| \,\hat r_i - \vec p_{(Z_i)}$.
So far we use only six constraints to reconstruct a three-momentum of $\vec p_{(F_i)}$. We can have two more constraints from our physics motivation where we assume the symmetric condition of $m_{F_1} = m_{F_2}$ and  $m_{\chi_1} = m_{\chi_2}$.  Thus we have eight constraints enough to measure $(m_F, m_\chi)$ even with one event. To illustrate how one can use Eq.\,(\ref{eq:func}), we take an event with $(m_F, m_\chi) = (200,50)\,\GeV$. In Fig.\,\ref{fig:solution}, each function for $m_{F_i}$ in Eq.\,(\ref{eq:func}) is presented either  a blue or a orange line with a crossing point as the solution for $(m_F, m_\chi)$.

\section{The HL-LHC study}
\label{sec:filter}
\subsection{Detector effects with limited number of events}

\begin{figure*}[t!]
\includegraphics[width=0.35\textwidth]{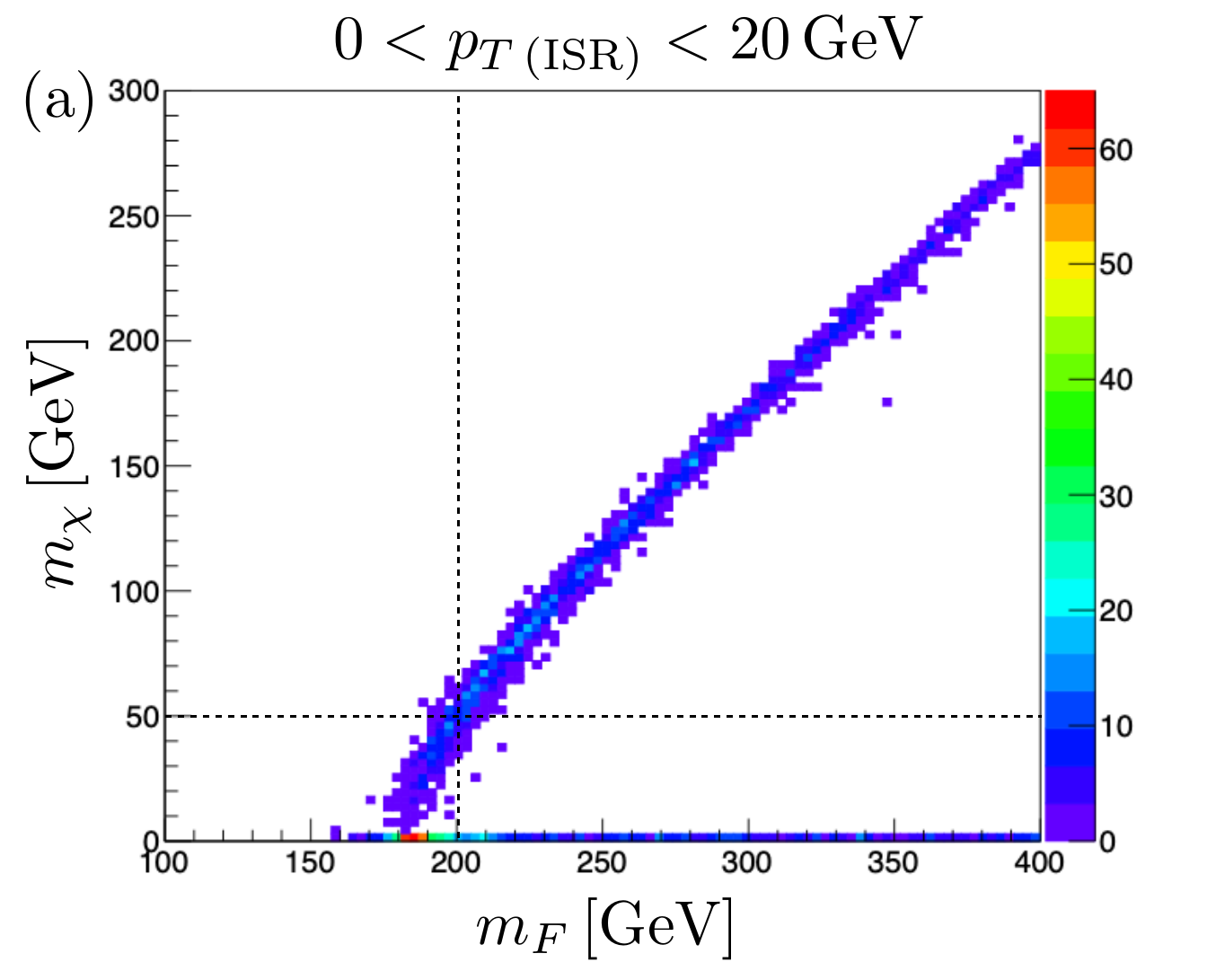}\qquad
\includegraphics[width=0.35\textwidth]{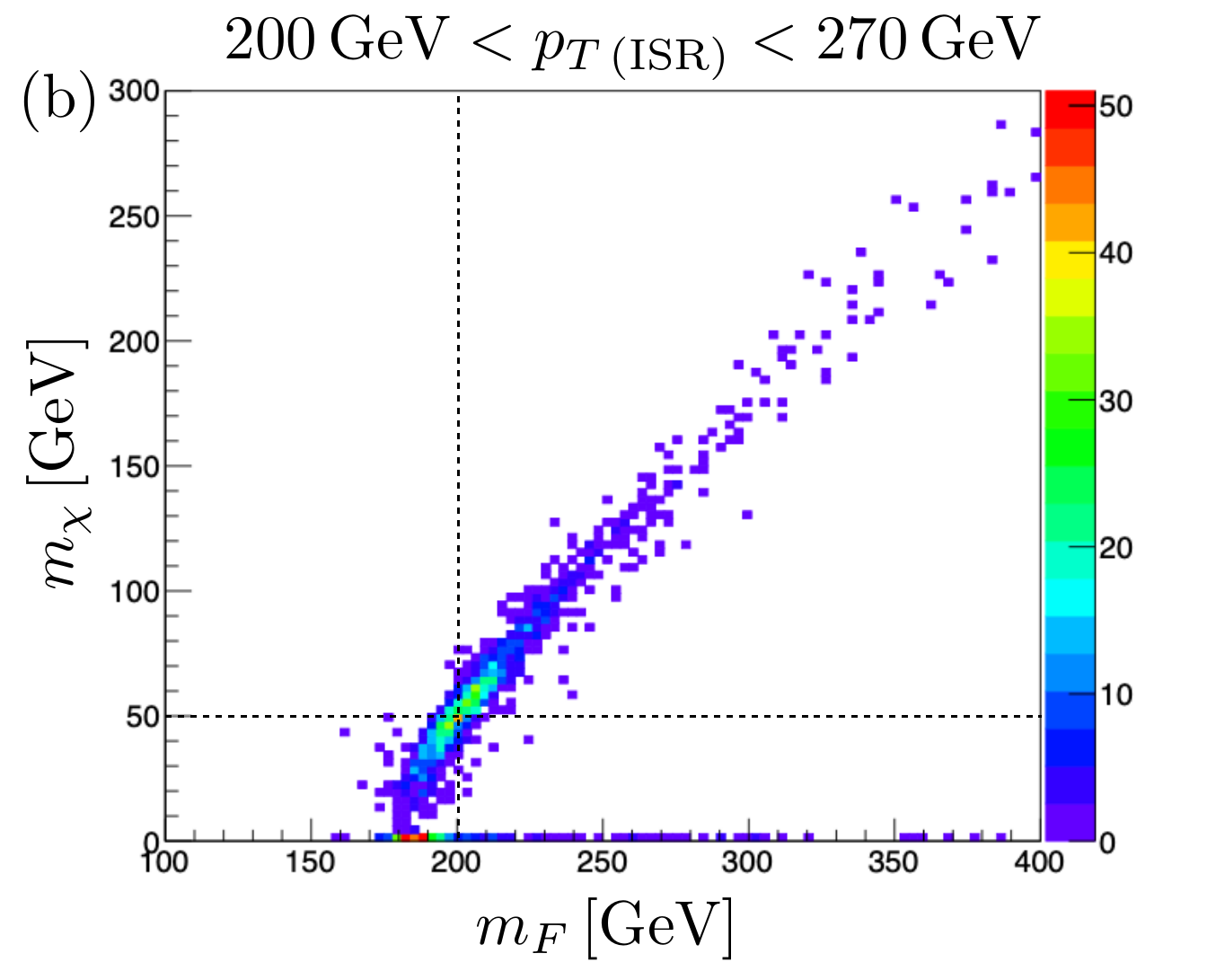}
\caption{Distribution of solutions in Eq.\,(\ref{eq:func}) for a mass spectrum $(m_F, \,m_\chi)= (200, \,50)\,\GeV$. We use 2000 events for each $p_{T(\textrm{ISR})}$ region using standard parton-level Monte Carlo simulations with Gaussian smearing for 
detector effects. }
\label{fig:ISR}
\end{figure*}
\begin{figure*}[t!]
\includegraphics[width=0.35\textwidth]{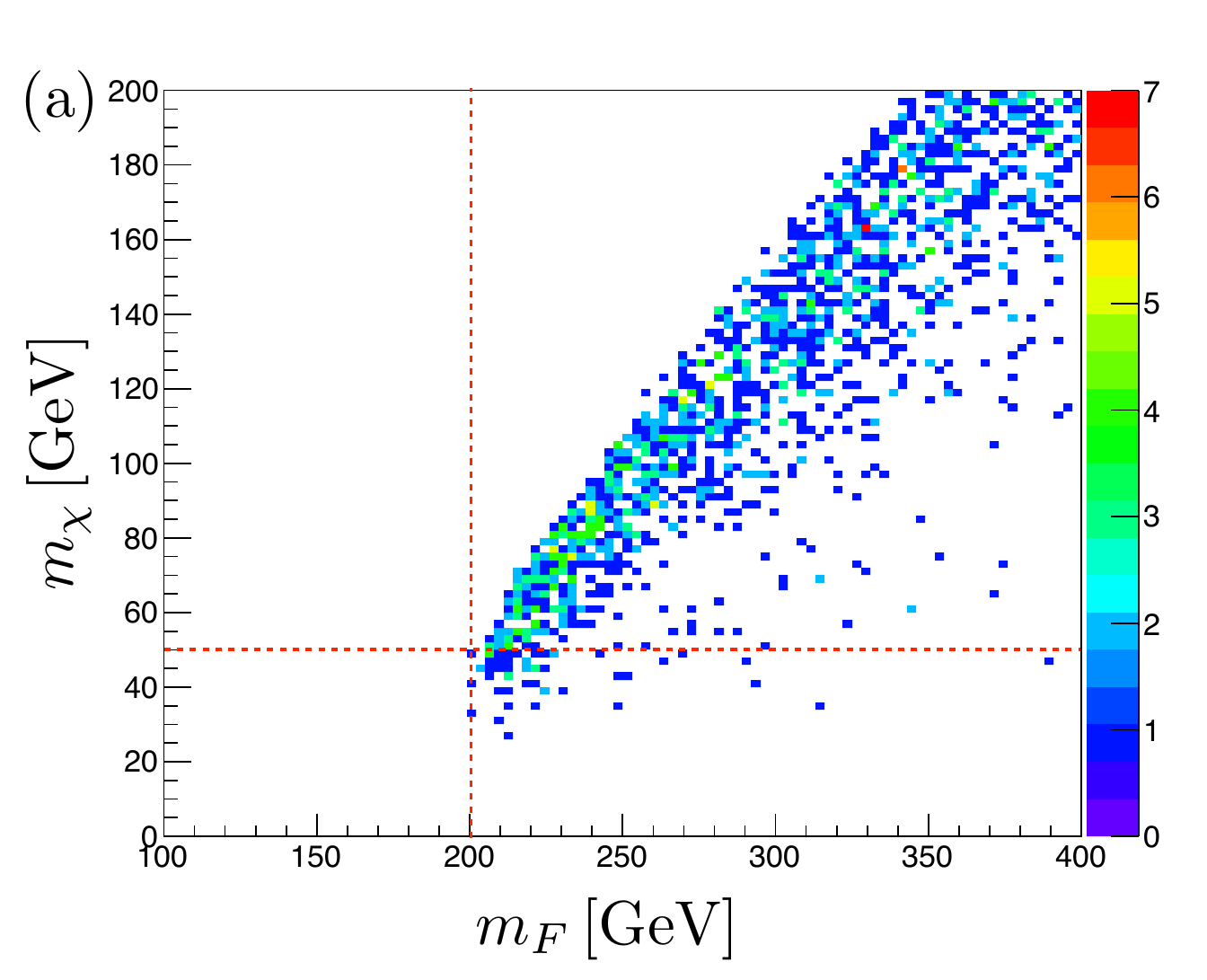}\qquad
\includegraphics[width=0.35\textwidth]{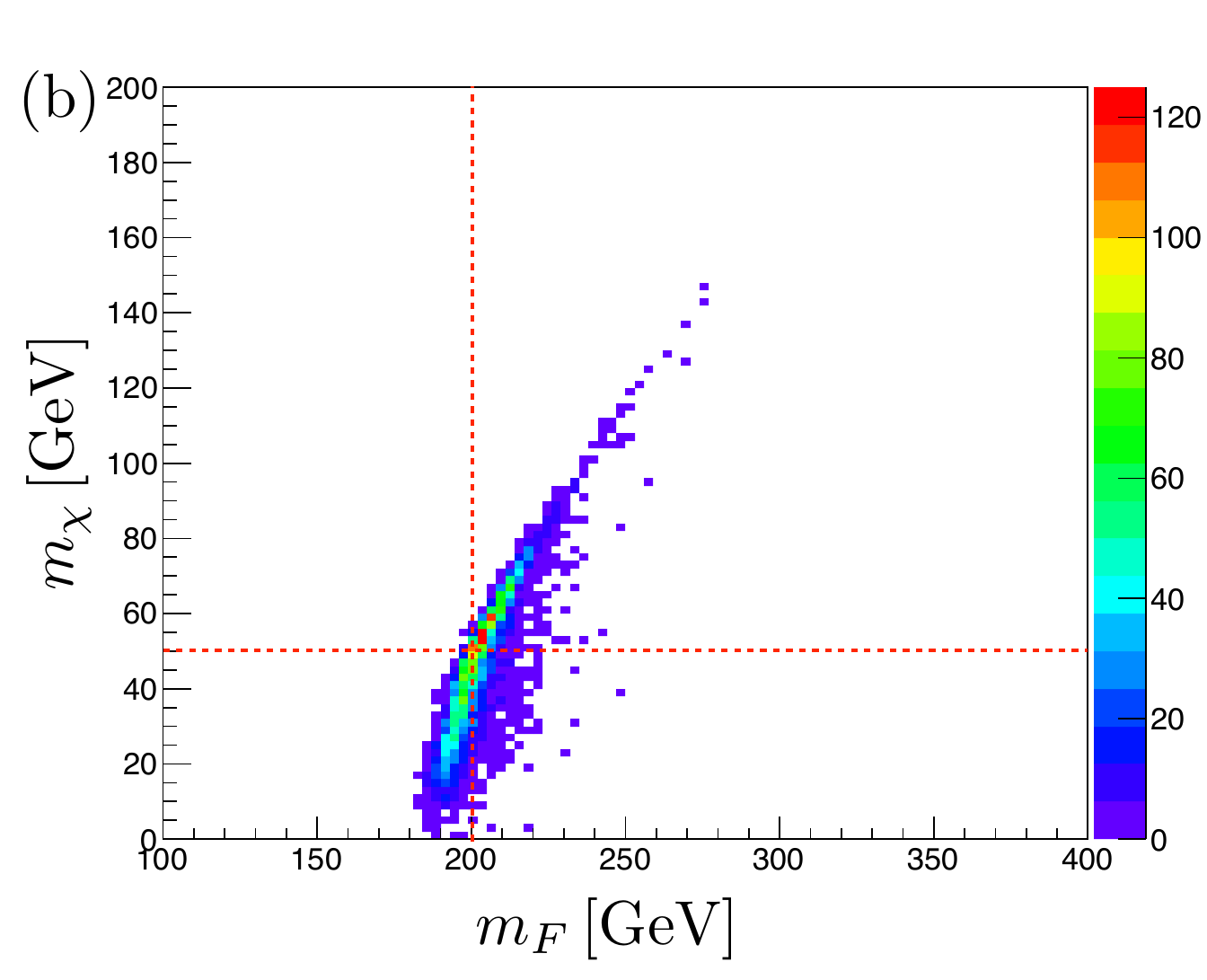}
\caption{We show the result of 5000 pseudo-experiments (a) without (b) with the filtering algorithm. We have only 20 events per one pseudo-experiment. For each experiment, we take the mean values for mass measurement. 
The mass spectrum is the same as in Fig\,\ref{fig:ISR}.
 }
\label{fig:filter}
\end{figure*}
We consider a channel where DVs can be marked by tracing the origin of $Z$ boson production using clean tracks of leptons 
 from $Z\to \ell^+ \ell^-$ for a precise measurement compared to hadronically decaying $Z$. However in the LHC detectors, there are uncertainties from (1) measurements of lepton momenta, which can be modeled with $2\%$ uncertainty\,\cite{Aad:2016jkr,Aaboud:2018ugz}, 
 (2) DV position resolution with a conservative value of $0.5\,\textrm{mm}$\,\cite{ATL-PHYS-PUB-2019-013}, and (3) the missing transverse energy $\met$ resolution of $10\%$ in the region of $\met>200\,\GeV$\,\cite{Aaboud:2018tkc}. These smearing effects will result in incorrect solutions for Eq.\,(\ref{eq:func}). 
 
 To numerically point this out, we simulate events with the study point of $(m_F, m_\chi) = (200,\,50)$\,GeV using Monte Carlo simulations by modeling smearing effects with Gaussian functions.
 As we discussed earlier, $p_{T(\textrm{ISR})}$ can enhance a numerical stability against smearing effects. We can observe this clearly in Fig.\,\ref{fig:ISR} where we divide Monte Carlo events according to the range of $p_{T(\textrm{ISR})}$. 
 In reality, however, we have $\mathcal{O}(10)$ events after cuts at the HL-LHC, so we cannot rely on the high $p_{T(\textrm{ISR})}$ configurations as most of events are located at low $p_{T(\textrm{ISR})}$. Thus we need to develop another method to reduce smearing effects. 

 To achieve this goal, we develop a simple filtering algorithm. Basically we rely on the fact that events are independent of one another. 
 Thus we focus on the clustering structure of solutions near the true mass point in $(m_F, \,m_\chi)$ solution-space;
\begin{itemize}
\item Each solution is treated as a vector $\vec{v}_i $ at $m_F-m_\chi$ plane with $\vec{v}_i = \left( m_{Fi} \ , \ m_{\chi i} \right)$.
\item For each $\vec{v}_i$, we measure an ``average distance'' $d_i$,
\beq
d_i = \frac{1}{N} \sum\limits^N_{j=1} |\vec{v}_i - \vec{v}_j|\, ,
\eeq
where $N$ is the total number of solution vectors,
\item Remove the $\vec{v}_i$ with largest $d_i$ from our vector list,
\item Calculate all $d_i$ again (only use the remaining vectors), and remove the vector with largest average distance. Repeat this process until only half of the vectors remain. 
\end{itemize}
This ``filtering'' algorithm is neatly dropping bad solutions in a simple and systematic way as in FIG.\,\ref{fig:filter}.
 
\subsection{Numerical studies with bechmark points}
\begin{table}[ht!]
\begin{center}\begin{tabular}{|c|c|c|c|c|}\hline B.\,P. & ~$m_F$\,[GeV]~ & ~$m_{\chi}$\,[GeV]~ & ~$ c\tau$\,(mm) & $N_{\textrm{events}}$@\,HL-LHC \\
\hline ~A~ & ~$200$  & ~$0$~ & $100$ & ~$30$~ \\
\hline B & $200$ & $108$ & $100$& $20$ \\
\hline C & $800$ & $0$ & $100$& $12$  \\
\hline D & $800$ & $708$ & $100$ & $24$  \\
\hline \end{tabular} \caption{Four benchmark points are illustrated. As the LHC detector is not sensitive to probe a mass scale less than $\mathcal{O}(1)\,\GeV$, we fix $m_\chi = 0$ for the mass scale of $\mathcal{O}(1-100)$\,keV.
$N_\textrm{events}$ is determined by recasting current LHC searches. These numbers are after baseline selection cuts.}
\label{tab:benchmark}
\end{center}
\end{table}

To illustrate our method with realistic situations at the HL-LHC, we consider four benchmark points divided by the scale of $m_F$ and a mass splitting of $\left\{m_F - (m_\chi+ m_Z)\right\} / m_F$ as in Tab.\,\ref{tab:benchmark}.
For benchmark points B and D, the mass splittings are highly suppressed, {\it i.e.}, $\left\{m_F - (m_\chi+ m_Z)\right\} / m_F \le \mathcal{O}(1) \% $ while mass splittings are more than $50\%$ in A and C. Thus if we measure 
mass spectrum $(m_F, m_\chi)$ precisely, we can identify where the lifetime of $m_F$ comes either from suppressed phase-space of small mass splitting or from very small coupling of $F$-$\chi$-$Z$. 
Here we consider the same lifetime of $F$ ($c\tau=100$~mm) to focus on kinematic differences and performance of our method which relies only on the reconstruction of $\vec p_{(F_i)}$, 
By recasting current long-lived particle search of ATLAS\,\cite{Aaboud:2017iio}, we expect 30, 20, 12 and 24 events, respectively for benchmark point A, B, C and D at the HL-LHC. 
We describe detailed information of recasting procedure in Appendix\,\ref{app:rec}.

For parton-level Monte Carlo simulations, we use \verb|MadGraph5|\,\cite{Alwall:2014hca} to generate events with ``MLM" jet matching algorithm\,\cite{Mangano:2006rw} implemented in \verb|Pythia8|\,\cite{Sjostrand:2007gs} for ISR jets. To cluster ISR jets, we use \verb|Fastjet|\,\cite{Cacciari:2011ma} 
with anti-$k_t$ algorithm\,\cite{Cacciari:2008gp}.
For detector effects, we consider detector geometries and smearing effects as we described above. 
Finally we choose 14 TeV collision energy and 3\,ab$^{-1}$ luminosity for the HL-LHC. 
As we need to enhance precision for DV, we focus only on leptonic channels of $Z$ decays. 
For baseline selection cuts, we require
\begin{itemize}
\item missing transverse energy $\met > 200\,\GeV$ to improve the $\met$ resolution~\cite{Aaboud:2018tkc}, 
\item 4 electrons or muons with $p_T>10\,\GeV$ in the final state,
\item 2 DVs to be reconstructed inside the inner detector ($4\,\textrm{mm}<L_{xy}<1\,\textrm{m}$ and $|L_{z}|<1\,\textrm{m}$)
where DVs are reconstructed by displaced tracks with impact parameter larger than 2~mm and $p_{T} >1\,\GeV$, 
two tracks of each DV to be matched to 2 leptons, and both DV mass $m_\text{DV}$ 
to be larger than $80\,\GeV$.\footnote{It is not likely that background processes have $m_\text{DV}>80\,\GeV$~\cite{Aaboud:2017iio}, 
so this cut leads background-free analyses.
}
\end{itemize}

Due to detector effects and limited statistics, the mass measurement based on track information has large uncertainties.
In order to improve the precision, we apply filtering algorithm.
It is worth mentioning a subtlety in solving Eq.\,(\ref{eq:func}) especially when a dark matter is very light. 
There are cases where two lines in FIG.\,\ref{fig:solution} do not cross due to smearing effects. As $m_{F_i}(m_{\chi_i})$ is an increasing function, we take $m_\chi =0$ and 
$m_F = \left( m_{F_1}(m_{\chi_1} = 0) + m_{F_2}(m_{\chi_2} = 0) \right)/2 $ for a solution. 

For benchmark points in Tab.\,\ref{tab:benchmark}, we perform 5000 pseudo-experiments to reduce statistical fluctuations. 
In Table~\ref{tab:result}, we tabulate 
the most probable mass values for each benchmark point and also estimated statistical errors of pseudo-experiments which are defined by the ``root-mean-square (RMS)'' value with respect to the most probable values
$
\text{RMS} = \sqrt{\sum^N_{i=1} (m_i - m^\text{peak})^2/\,N}\, ,
$
where $m^\text{peak}$ is the most probable value and $N$ is the number of pseudo experiments.
\begin{table}[t!]
\begin{center}\begin{tabular}{|c|c|c|}\hline 
\multirow{2}{*}{B.\,P.}&  $(m_F,\, m_\chi)^\textrm{true}$ &  \multirow{2}{*}{  RMS  }\\
\cline{2-2}
& $(m_F,\, m_\chi)^\textrm{peak}$  &  \\ 
\hline
\multirow{2}{*}{A}&  $(200, \, 0)$ &  \multirow{2}{*}{ $(7.3,\,13)$ } \\
\cline{2-2}
&  $(208,\,14)$ &  \\ 
\hline
\multirow{2}{*}{B}&  $(200, \, 108)$ &  \multirow{2}{*}{ $(14,\, 15)$ } \\
\cline{2-2}
&  $(199,\,105)$ &  \\
\hline
\multirow{2}{*}{C}&  $(800, \, 0)$ &  \multirow{2}{*}{ $(250,\, 360)$ } \\
\cline{2-2}
&  $(835,\,0)$ &  \\
\hline
\multirow{2}{*}{D}&  $(800, \, 708)$ &  \multirow{2}{*}{ $(79,\, 80)$ } \\
\cline{2-2}
&  $(792,\,694)$ &  \\
\hline \end{tabular} \caption{Peak measured values and root-mean-squares (RMS) for four benchmark points. Numbers are in GeV unit. }
\label{tab:result}
\end{center}
\end{table}
For benchmark points A, B, and D, the results show good precision, where the errors are around 10\% of mother particle mass.
For benchmark point C, on the other hand, RMS error is around $30-45$\% of mother particle mass.
This mainly comes from small statistics as C has only 12 events after cuts which has factor 1/2 - 1/3 reduction compared to other benchmark points.

In the following, we discuss more direct implication of the above analyses in the freeze-in dark matter scenarios.
As argued in Ref.~\cite{Bae:2017dpt}, as the phase space of a mother particle decay is highly suppressed, the distribution of DM produced from this decay process becomes colder.
Meanwhile, the distribution of DM produced by 2-to-2 scattering processes remains almost the same.
This feature can be probed by observations such as Ly-$\alpha$ forest data if the DM mass is of order keV.
Therefore, by observing mass spectrum of DM sector at the LHC, we can infer the relative ``warmness'' of DM distribution.
In order to illustrate this situation, we introduce an additional benchmark point E, on which $(m_F, \,m_\chi)=(100,\,0)\,\GeV$,  $c\tau=100$~mm. The expected number of events after baseline selection cuts at the HL-LHC is $10$.
At the LHC, we measure $(m_F, \,m_\chi)^\textrm{peak} = (101,\,0)\,\GeV$ with statistical deviation RMS = $(81, 47)\,\GeV$ from 5K pseudo experiments. 
Thus, even though we have uncertainties in determining mass spectrum due to the limited statistics, we will have hints of dark matter temperature of our universe from the HL-LHC. 

\section{Conclusions}
\label{sec:con}
We develop a pure kinematic method to determine mass spectrum involving DVs at the LHC by locating visible particle track inside inner tracking detector. To measure mass spectrum, we assume conditions $(m_{F_1} = m_{F_2})$ and $(m_{\chi_1} = m_{\chi_2})$ for both decay chains. We also require ISR jets not to have back-to-back configuration between $F$'s which results in a null solution in reconstructing three momentum vectors of $F$'s. Large ISR can enhance a numerical stability for the determination of mass spectrum, but we would not have enough statistics to focus on the large $p_{T(\textrm{ISR})}$ region at the HL-LHC. 

 In this study, we consider only $\mathcal{O}(10)$ events with detector smearing effects. The performance of the mass measurements gets degraded due to imprecise information on $\met$ and four vectors of leptons. To achieve satisfactory performance with small number of events, we propose a simple and systematic method which removes solutions in the mass reconstruction with large errors. 

We have demonstrated that one can achieve within $\mathcal{O}(10)\%$-level precision on mass measurements at the HL-LHC. This result from the HL-LHC enables us to understand the origin of DV signals, either due to a suppressed phase space or due to a feeble coupling of decaying particle. In this respect we can see a relationship between the freeze-in mechanism in the early universe and DV signature at the LHC.

Finally, as we do not rely on any new features of upcoming detector upgrades, our proposed method is orthogonal to other studies with utilizing timing information\,\cite{Liu:2018wte,Kang:2019ukr}.  
One can enhance a precision in mass measurements by combining results from both methods once the LHC identifies DV signals from a new physics. 

\acknowledgments{This work was supported by IBS under the project code IBS-R018-D1. MP is supported by Basic Science Research Program through the National Research Foundation of Korea Research Grant NRF-2018R1C1B6006572. 
MZ is supported by the National Natural Science Foundation of China (Grant No. 11947118). 
This work was performed in part at Aspen Center for Physics, which is supported by National Science Foundation grant PHY-1607611.
}

\appendix

\section{Recasting of Current long-lived particle searches }
\label{app:rec}

\begin{figure*}[ht!]
\includegraphics[width=0.35\textwidth]{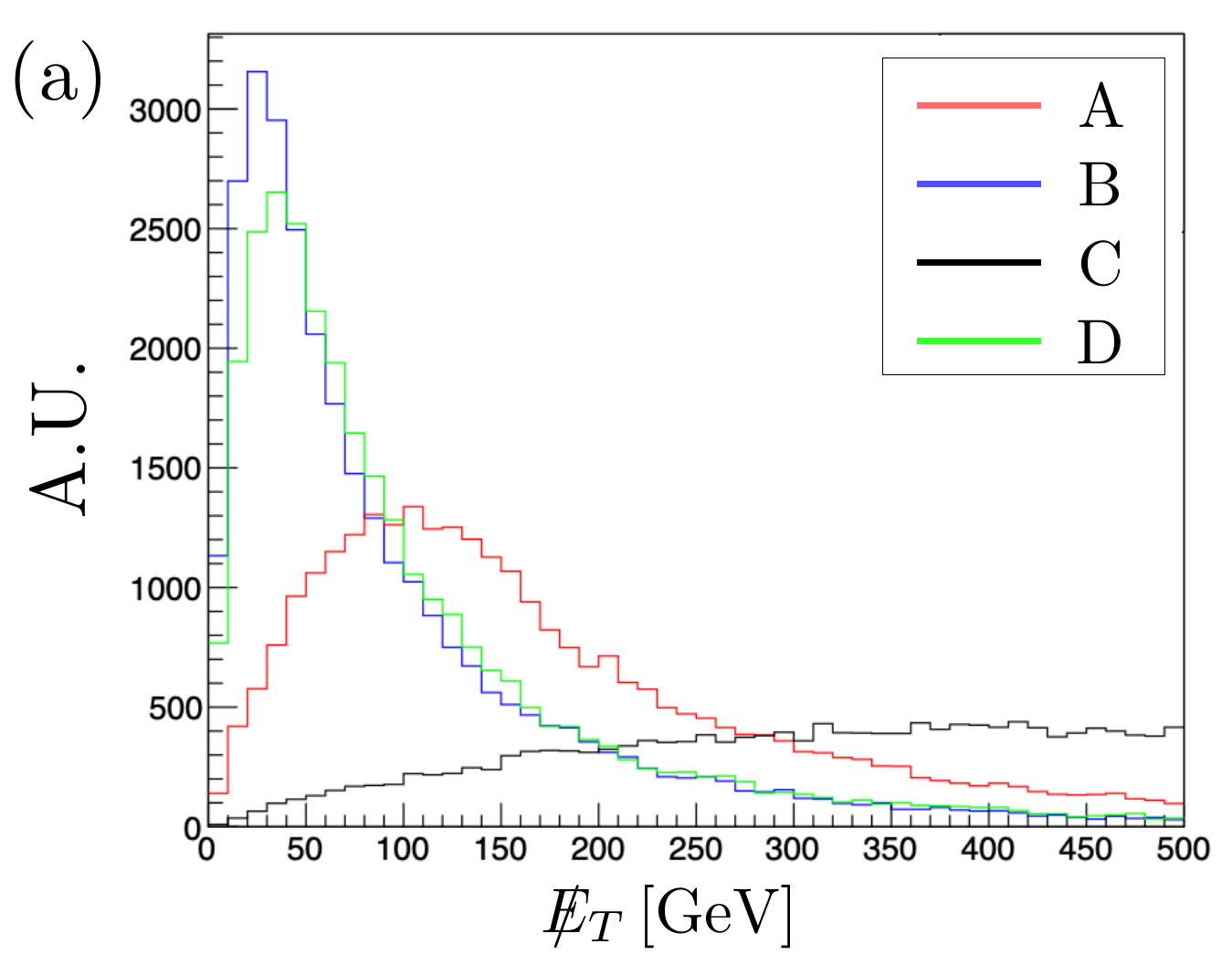}\qquad
\includegraphics[width=0.35\textwidth]{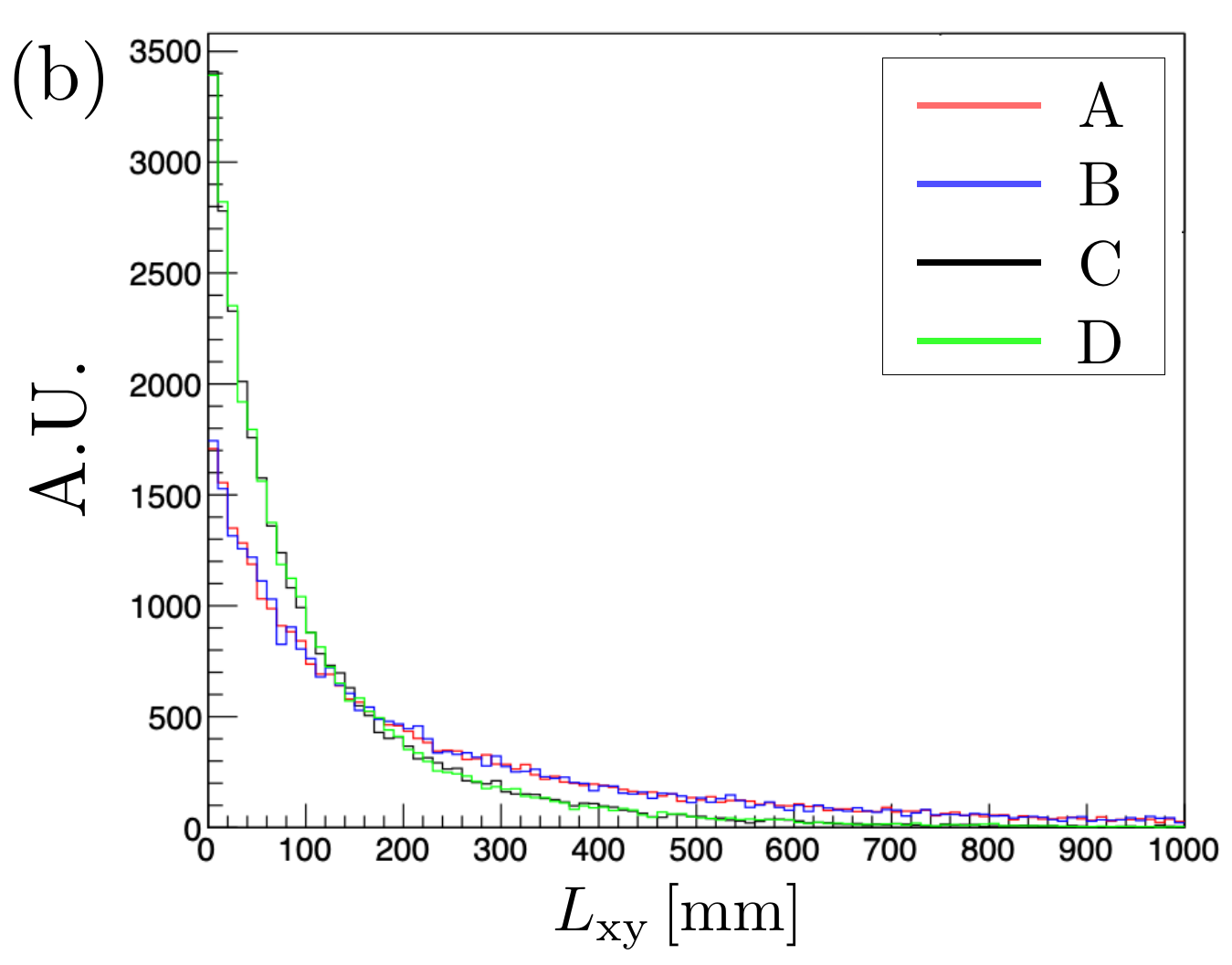}
\caption{Distributions of $\met$ and $L_{\textrm{xy}}$ for benchmark points. Benchmark points B and D with small mass splittings have lower $\met$ distributions than those of A and C.
For $L_{\textrm{xy}}$, C and D with larger $m_F$ have smaller values than the case of A and B. $N_\textrm{tracks}$ and $m_\textrm{DV}$ distributions are not significantly different among benchmark points.}
\label{fig:effi}
\end{figure*}
We recast the ATLAS long-lived massive particles search report\,\cite{Aaboud:2017iio} to estimate the upper limit of $\sigma(pp\to FF)$.
There are another long-lived particle searches, for example, searches with displaced jets in the ATLAS\,\cite{Aaboud:2019opc}, which do not have sensitivity for our scenario as 
they need information of calorimeters. 

Here we briefly describe cut-flow used in the ATLAS report\,\cite{Aaboud:2017iio}.
\begin{itemize}
\item Transverse missing energy $\met \ge 250\,\GeV$.
\item 75\% of events need to have at least one trackless jet with $p_\text{T} >70\,\GeV$ or at least two trackless jets with $p_\text{T} > 25\,\GeV$. For other 25\% events there is no requirement on trackless jet. 
Trackless jet is defined to be a jet with $\sum p^\text{track}_\text{T}< 5\,\GeV$. 
\item At least one DV needs to be reconstructed in fiducial region of inner detector within a transverse position $4\,\textrm{mm}< L_\text{xy}<300\,\textrm{mm}$ and longitudinal position $|L_\text{z}|<300\,\textrm{mm}$. DV is reconstructed by displaced tracks with impact parameter larger than 2\,mm and $p_\text{T} > 1\,\GeV$. More than five displaced tracks ($N_\text{tracks}$) from a DV are required, and reconstructed mass with tracks from DV ($m_\text{DV}$) needs to be larger than $10\,\GeV$. 
\item 42\% fiducial volume need to be discarded due to huge backgrounds from hadronic interaction in material rich region. 
\end{itemize}

DV tagging efficiency is a function of its position, $N_\text{tracks}$, and $m_\text{DV}$. 
As suggested in\,\cite{Brooijmans:2018xbu}, we perform our DV tagging by utilizing detailed information provided by ATLAS group in the auxiliary information of\,\cite{Aaboud:2017iio}. 
Acceptances of our 4 benchmark points after performing above selection criteria are listed in Tab.\,\ref{accep}.
Acceptances for compressed spectrum B and D are much smaller than large mass splitting spectrum A and C as compressed spectrum does not provide large enough $\met$ as in Fig.\,\ref{fig:effi}.

\begin{table}[htp]
\begin{center}\begin{tabular}{|c|c|c|c|}\hline B.\,P.  & Acceptance & $\sigma(pp\to FF)^{\rm max}_{\rm 13~TeV}$ & $\sigma(pp\to FF)^{\rm max}_{\rm 14~TeV}$  \\
\hline A & 0.63\%   &  14.43  & 17.56   \\
\hline B & 0.14\%  & 64.41  & 78.41  \\
\hline C & 6.76\%  & 1.35  & 1.87  \\
\hline D &  0.52\%  & 17.62  & 24.34  \\
\hline \end{tabular} \caption{Acceptance, upper limits on production cross-section at 13TeV, and recasted cross-section at 14TeV LHC. Values are in fb unit.  }
\label{accep}
\end{center}
\end{table}

\bibliographystyle{utphys}
\bibliography{filhc}

\end{document}